\documentclass[12pt]{iopart}
\usepackage{graphicx}
\usepackage{dcolumn}
\usepackage{bm}

\begin{document}

\title{Torsion-balance tests of the weak equivalence principle}

\author{T A Wagner, S Schlamminger\footnote{current address: National Institute of Standards and Technology, Gaithersburg,
Maryland 20899, USA}, J H Gundlach  and E G Adelberger}
\address{Center for Experimental Nuclear Physics and Astrophysics, Box 354290,
University of
Washington, Seattle, WA 98195-4290}
\ead{eric@npl.washington.edu}

\begin{abstract}
We briefly summarize motivations for testing the weak equivalence principle and then review recent torsion-balance results that compare the differential accelerations of beryllium-aluminum and beryllium-titanium test body pairs with precisions at the part in $10^{13}$ level. We discuss some implications of these results for the gravitational properties of antimatter and dark matter, and speculate about the prospects for further improvements in experimental sensitivity.
\end{abstract}
\submitto{\CQG}
\maketitle
\setlength{\floatsep}{0.25in}

\section{General framework}
The weak equivalence principle (WEP) states that in a uniform gravitational field all objects, regardless of their composition, fall with precisely the same acceleration. In Newtonian terms, the principle asserts the exact identity of inertial mass $m_i$ (the mass appearing in Newton's second law) and gravitational mass $m_g$ (the mass appearing in Newton's law of gravity). The WEP implicitly assumes that the falling objects are bound by non-gravitational forces. The strong equivalence principle extends the universality of free fall to objects (such as astronomical bodies) where the effects of gravitational binding energy cannot be neglected. 

WEP tests were traditionally interpreted in Newtonian terms, {\em i.e.} as searches for possible departures from
exact equality of $m_g/m_i$ for objects 1 and 2 as specified by the E\"otv\"os parameter
\begin{equation}
\eta_{1,2}=\frac{a_1 -a_2}{(a_1 +a_2)/2}=\frac{(m_g/m_i)_1 -(m_g/m_i)_2}{[(m_g/m_i)_1 +(m_g/m_i)_2\:]/2}~,
\end{equation}
where $a$ is the measured free-fall acceleration.
In this case, the properties and location of the attractor toward which the objects were falling was irrelevant. For technical reasons described below, the classic experiments at Princeton\cite{ro:64} and Moscow\cite{br:71} used the sun as the attractor.

However, as emphasized by Fischbach\cite{fi:86}, it is appropriate to view WEP tests as probes for possible new Yukawa interactions, potentially much weaker than gravity, that would be essentially undetectable by other means. In this case, we ascribe any violation of the WEP to a previously unknown Yukawa interaction arising from quantum exchange of new bosons that couple to vector or scalar charges of the test bodies and attractor. Vector or scalar boson exchange forces of quantum field theories produce a spin-independent potential between test body $i$ and attractor $A$
of the form
\begin{equation}
V_{\rm OBE}(r)=\mp \frac {\tilde{g}^2}{4\pi}\frac{\tilde{q}_i \tilde{q}_A}{r}\exp({-r/\lambda})~,
\end{equation}
where $\tilde{q}$ is a fermion's scalar or vector dimensionless charge, $\tilde{g}$ is a coupling constant, and 
$\lambda=\hbar/(m_b c)$ is the range of the force mediated by bosons of mass $m_b$. The $-$ and $+$ signs apply to scalar and vector interactions, respectively. The total potential can be written in a form appropriate for WEP tests as
\begin{equation}
V_{i,A}=V_{\rm G}+V_{\rm OBE}=V_{\rm G}(r)\left[ 1+\tilde{\alpha}\left(\frac{\tilde{q}}{\mu}\right)_i\left(\frac{\tilde{q}}{\mu}\right)_A\exp({-r/\lambda}) \right]~,
\end{equation}
where the dimensionless ratio $\left(\tilde{q}/\mu\right)$ is an object's charge per atomic mass unit ($u$), and 
the dimensionless Yukawa strength parameter
\begin{equation}
\tilde{\alpha}=\pm\tilde{g}^2/(4 \pi G u^2)~.
\end{equation}
In this case
\begin{equation}
\eta_{1,2}=\tilde{\alpha}\:\left[\left(\frac{\tilde{q}}{\mu}\right)_1\!-\left(\frac{\tilde{q}}{\mu}\right)_2 \right]\:\left(\frac{\tilde{q}}{\mu}\right)_A \left(1+\frac{r}{\lambda}\right)\exp{(-r/\lambda)}~.
\end{equation}
For electrically neutral bodies consisting of atoms with proton and neutron numbers $Z$ and $N$, respectively, a general vector charge-to-mass ratio can be parameterized as
\begin{equation}
(\tilde{q}/\mu)=(Z/\mu)\cos\tilde{\psi}+(N/\mu)\sin \tilde{\psi}~~~{\rm with}~~~
\tan\tilde{\psi}\equiv \frac{\tilde{q}_n}{\tilde{q}_e+\tilde{q}_p}~,
\label{eq: charge}
\end{equation}
where $\tilde{\psi}$ is an unknown parameter that ranges between $-\pi/2$ and $+\pi/2$. 
It is easy to see that any vector interaction must violate the equivalence principle because particles and anti-particles have opposite vector charges. Less dramatically, different atoms composed of ordinary matter must also have different vector $(\tilde{q}/\mu)$ ratios because the vector charge of an atom is the sum of the charges of its ingredients, but the atom's mass is less than the mass of its ingredients because of binding energy. 

Note that the $\tilde{q}$ of any given atom vanishes for some value of $\tilde{\psi}$. This implies that to make a comprehensive and unbiased test of the WEP it is necessary to
\begin{enumerate}
\item{test with 2 different test-body composition dipoles falling toward 2 different attractors to avoid accidental cancellations of the charges of the test body dipole or the attractor, and}
\item{use attractors with the smallest practical distance from the test bodies to cover a wide span of Yukawa ranges $\lambda$.}
\end{enumerate}

The situation for scalar charges is considerably more complicated because scalar charges are neither conserved nor Lorentz invariant (the charge density rather than the charge itself is a Lorentz scalar) and binding energy along with virtual fermion-antifermion loops carry scalar charges. Furthermore, a scalar interaction can couple to $T_{\mu}^{\mu}$, the trace of the energy-momentum tensor {\em i.e.} effectively to mass. (Note that, for massless scalar fields, the $T_{\mu}^{\mu}$ term has no experimental significance because it would indistinguishable from normal gravity.) Therefore,  detailed field-theoretic calculations are required to compute scalar charges of neutral atoms.  Nevertheless, one expects the WEP component of a general scalar charge-to-mass ratio 
will be roughly described by equation \ref{eq: charge} as well. 

A particularly interesting scalar interaction arises from dilaton exchange, where the dilaton is the scalar partner of the massless graviton that is inherent in string theories. Kaplan and Wise\cite{ka:00} found that 
the force generated by low-mass dilatons is dominated by coupling to the gluon field strength, which gives a force between nucleons that is $\sim 10^3$ times stronger than gravity with only a small, 0.3\%, WEP-violating component. A long-range dilaton field with such couplings is clearly ruled out by many experiments. It is, therefore, usually assumed that the dilaton has a finite mass which gives its force a small range, allowing it to evade the experimental bounds. We will return to this issue in  Sec.~\ref{subsec:implications}. Donoghue and Damour\cite{do:10,do:10a} made related calculations 
but took an entirely different point of view.
They computed the WEP-violating effects but allowed the various dilaton-coupling terms to be free parameters, but implicitly  assumed that composition-independent coupling to the gluon field was dominant. They suggested  that WEP experiments be analyzed to set bounds on bilinear combinations of $d_g$, $d_e$ (couplings to gluon and electromagnetic field strength) and $d_{m_e}$, $d_{\hat m}$ and $d_{\delta m}$ (couplings to the masses of the electron and to the average and difference of the up and down quark masses). The dominant WEP-violating effects are expected to arise from $d_e$ and $d_{\hat m}$.

\section{Principles of torsion-balance tests of the WEP}
The remarkable sensitivity of torsion-balance null tests of the WEP results from two properties of such instruments:
%
%
\begin{figure}[b]
\begin{center}
\includegraphics[width=0.4\columnwidth]{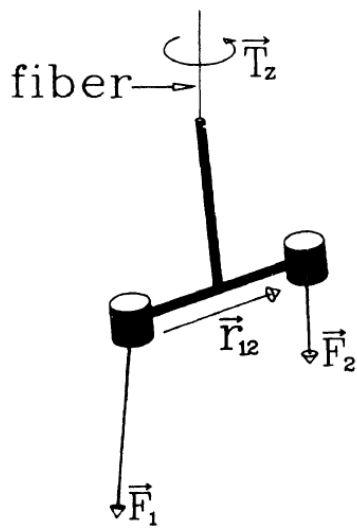}
\end{center}
\caption{Operating principle of the E\"otv\"os torsion balance. This idealized balance consists of two test bodies attached to a rigid, massless frame that is supported by a perfectly flexible torsion fibre. ${\bm F_1}$ and ${\bm F_2}$ denote the external forces on the test bodies. The torque about the fibre axis is $T_z=({\bm F_1} \times {\bm F_2}\cdot {\bm r_{12}})
/|{\bm F_1}+{\bm F_2}|$. The signal is the change in $T_z$ when the instrument is rotated about the fibre axis so that the component of $\bm r_{12}$ along the direction of ${\bm F_1}\times{\bm F_2}$ changes sign.
\label{fig: principle}}
\end{figure}
\begin{enumerate}
\item{A freely hanging torsion balance responds only to a {\em difference in the directions} of the external force vectors on the test bodies and not on their magnitudes\cite{ad:90} (see figure~\ref{fig: principle}). This allows instruments with tolerances at the $10^{-5}$ level to make measurements with a precision of a part in $10^{13}$. In fact, current experiments are limited by gravity gradients which, when coupled to imperfections in the geometry of the torsion pendulum, also give a difference in the {\em directions} of the forces on the test bodies\cite{su:94}. The highest precisions have been obtained by uniformly rotating the balance with respect to the attractor, giving a WEP-violating signal that is a sinusiodal function of the rotation angle. The classic WEP tests of the Princeton\cite{ro:64} and Moscow\cite{br:71} groups employed the sun as the attractor and let the earth's rotation provide the smooth rotation of the instrument.}
\item{Although an actual torsion oscillator has many modes (twist, pendulum, bounce, wobble, {\em etc.}) the frequency of the twist mode ($\sim$mHz) lies well below that of all the other modes ($\sim$Hz). Therefore the other modes can be damped before their energy has had much chance to leak into the twist mode, so that the torsion ocillator effectively has a single mode that can operate close to the thermal limit. 
The thermal torque noise at frequency $f$ in a single-mode torsional oscillator with fibre torsional constant $\kappa$ and quality factor $Q$ has a power spectral density (see ~\cite{sa:90})
\begin{equation}
\tau (f)^2 = 4 k_B T \kappa / (2 \pi f Q)~;
\label{eq: noise}
\end{equation}
where it is assumed that the damping is dominated by internal friction in the suspension fibre. (The related case of a rotating 2-dimensional oscillator is discussed in ~\cite{pe:11}.)}
\end{enumerate}
\section{Modern experimental tests and their results}
The instrument rotation scheme of the classic Princeton and Moscow experiments, while very smooth, had two main disadvantages:
\begin{enumerate}
\item{The 24 hour signal period posed serious problems. Most noise sources increase as the frequency decreases (as $1/f$ for fibre damping and $1/f^2$ for several other sources). Furthermore many possible systematic effects have a 24 h period (temperature, vibration, power fluctuations, etc.).}
\item{The solar attractor rendered the experiments completely insensitive to Yukawa forces with ranges less than $10^{11}$ m}. 
\end{enumerate}
To avoid these limitations, the E\"ot-Wash group developed a series of torsion balances equipped with uniformly-rotating turntables\cite{ad:90,su:94,sm:00,he:08,sc:08,wa:12}. 
This allowed the earth to be used as the attractor and placed the signal at the turntable's rotation frequency ($\sim$mHz for our apparatus). The centrifugal force due to earth's rotation pushes a torsion pendulum in the Northern Hemisphere toward the south. This force is balanced against a horizontal component of gravity, which in Seattle, Washington at a latitude of $47.7^{\circ}$N is 1.68 cm~s$^{-2}$, giving a maximum horizontal acceleration three times greater than that toward the sun.

The requirements 
on the constancy of the turntable rotation rate\cite{su:94}, as well as the alignment of its rotation axis with the suspension fibre, are quite severe. Suppose that the turntable rotation rate $\omega_{\rm tt}$ is not completely constant so that
\begin{equation}
\omega_{\rm tt}(t)=\omega_{\rm c}+\sum_{n=1}^N \omega_n e^{-in\omega_{\rm c}t}~.
\end{equation} 
This will induce a twist angle $\theta$ of the torsion pendulum
\begin{equation}
\theta(t)=\sum_{n=1}^{N} \frac{-in\omega_{\rm c}}{\omega_0^2-(n\omega_{\rm c})^2}\omega_n e^{-in\omega_{\rm c}t}~,
\label{eq: false effect}
\end{equation} 
where pendulum damping has been neglected and $\omega_0$ is the frequency of free torsional oscillations. The $n=1$ term will generate a spurious WEP signal that must be cancelled by combining data with 2 opposite orientations of the composition dipole in the rotating balance. 

The choice of $\omega_{\rm tt}$ involves competing considerations of thermal and other low-frequency torque noises
and the noise in the twist readout system. The response of a damped torsion oscillator to a torque of magnitude $T$  varying at a frequency $\omega_{\rm s}$ is
\begin{equation}
\theta(\omega_{\rm s})= \frac{T}{\kappa}\frac{\omega_0^2}{\sqrt{(\omega_0^2-\omega_{\rm s}^2)^2+(\omega_0^2/Q)^2}}~,
\end{equation}
where $\kappa$ is the torsional spring constant. Running on resonance ($\omega_{\rm s}=\omega_0$) is only sensible when the $\theta$ readout noise is completely dominant. Otherwise, the signal-to-noise ratio is optimized by a compromise between the thermal torque noise (which falls with increasing $\omega_{\rm s}$) and noise from imperfect turntable rotation (which rises as $\omega_{\rm s}$ increases).
%
%
\begin{figure}[hb]
\begin{center}
\includegraphics[width=0.6\columnwidth]{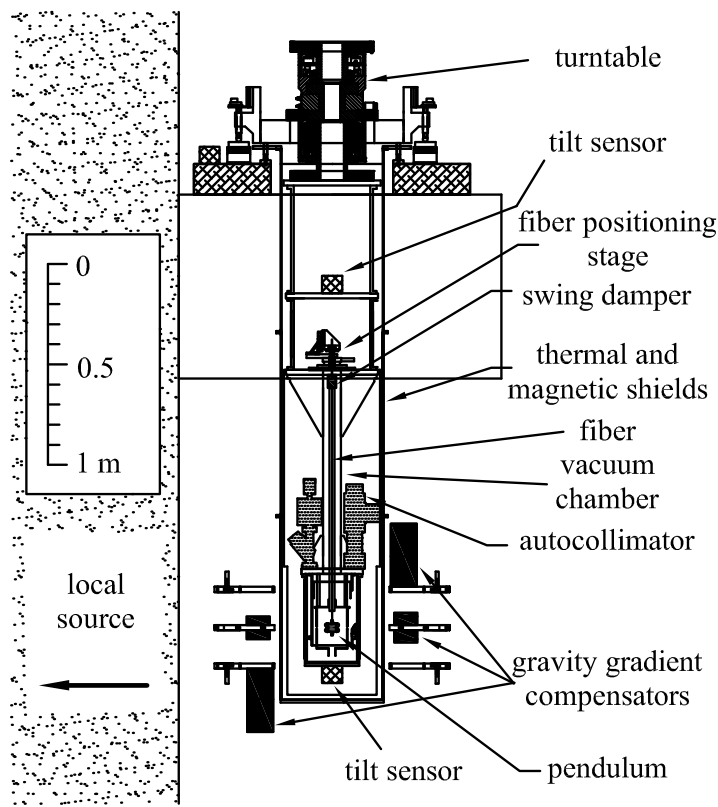}
\end{center}
\caption{Simplified scale drawing of the E\"ot-Wash WEP torsion balance.
\label{fig:apparatus}}
\end{figure}
The rotating torsion balance used for the recent E\"ot-Wash test of the WEP\cite{sc:08,wa:12} is depicted in figure \ref{fig:apparatus}. An air-bearing turntable driven by an eddy-current motor provided a highly uniform rotation rate. A laser autocollimator measured the twist of the torsion pendulum\cite{ho:04}. Additional sensors on the apparatus measured temperature, vacuum pressure, and tilts. Feedback to the tilt sensors aligned the rotation axis with local vertical by controlling thermal-expansion legs that supported the turntable\cite{he:08}. The balance was surrounded by passive thermal and magnetic shields. Large masses placed nearby compensated the leading static environmental gravity gradients by more than two orders of magnitude. An ion pump maintained the vacuum chamber at a pressure of $< 10^{-4} ~\mbox{Pa}$. The apparatus is located within a temperature-stabilized foam box inside a temperature-controlled room. 
The pendulum's twist angle and 27 other environmental sensors were recorded every $\approx 3 \mbox{s}$ by a data acquisition system. The recorded twist angle was passed through a digital notch filter to remove the pendulum oscillation, then separated into Fourier components by fitting the time series from two complete turntable rotations with sines and cosines of  harmonics of the turntable angle, plus a 2nd-order polynomial drift. 
%
\begin{figure}[ht]
\begin{center}
\includegraphics[width=0.35\columnwidth]{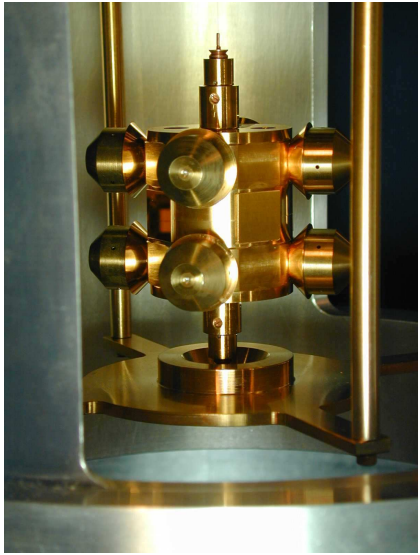}
\end{center}
\caption{[Colour online] Torsion pendulum used in the recent E\"ot-Wash WEP test. An Al frame holds 4 mirrors and supports 8 barrel-shaped test bodies, 
4 of which are Be and 4 are Ti or Al. The structure underneath the pendulum allows the pendulum to be parked to prevent damage when the apparatus is serviced and catches the pendulum if a small earthquake should break the suspension fibre. The tungsten fibre is just visible at the top.} 
\label{fig:pendulum}
\end{figure}

The torsion pendulum used for measurements with Be-Ti and Be-Al test body pairs, shown in Figure \ref{fig:pendulum}, was supported by a 1.07 m long, 20 $\mu$m thick tungsten fibre. The pendulum's design,  with 4-fold azimuthal and up-down symmetries, reduces systematic effects by minimizing the coupling to gravity gradients and by allowing for four different orientations of the pendulum with respect to the turntable rotor.
The gravitational multipole framework described in  \cite{su:94} was used to suppress couplings to environmental gravity gradient fields that fall off more slowly than $r^{-6}$, with the exception of the primary four-fold symmetry of the pendulum that gave a weak signal at the fourth harmonic of the turntable rotation frequency. This was readily distinguished from a WEP-violation whose signal is at the turntable rotation frequency. 

The test bodies, which comprise 40 g of the pendulum's 70 g mass, all have identical masses and outside dimensions to suppress systematic effects. They are removable, which allowed us to use two different composition dipoles and to rearrange test bodies to invert the composition dipole on the pendulum frame. This last strategy canceled systematic effects that followed the pendulum frame rather than the test bodies themselves.
The pendulum is coated with $\approx 300~\mbox{nm}$ of gold and is surrounded by a gold-coated electrostatic shield to minimize electrical effects from work-function variations. 
The test-body materials were selected for their scientific impact and for practical concerns such as mechanical stability and freedom from magnetic impurities. 
Table \ref{tab:charge} summarizes the charges of some test body materials.
\begin{table}
\caption{\label{tab:charge}Charge-to-mass ratios of selected test
body materials. $Z$, $N$ and $B$ are the atomic, neutron and baryon
numbers, respectively. $Q_{\hat{m}}$ and $Q_{e}$ are the dilaton
charge-to-mass ratios associated with the average light quark mass and
the electrostatic field strength, respectively \cite{do:10a}.}
\begin{indented}
\item[]\begin{tabular}{l cccc}
\br
        & Be & Ti & Al & Pt \\
\mr
$Z/\mu$         & 0.44384 & 0.45961 & 0.48181 & 0.39983  \\
$N/\mu$         & 0.55480 & 0.54147 & 0.51887 & 0.60032  \\
$B/\mu$         & 0.99865 & 1.00107 & 1.00068 & 1.00015  \\
$Q_{\hat{m}}$	& 0.07526 & 0.08267 & 0.08076 & 0.08526 \\
$Q_{e}$		& 0.00072 & 0.00228 & 0.00174 & 0.00428 \\
\br
\end{tabular}
\end{indented}
\end{table}

\begin{figure}[b]
\begin{center}
\includegraphics[width=0.9\columnwidth]{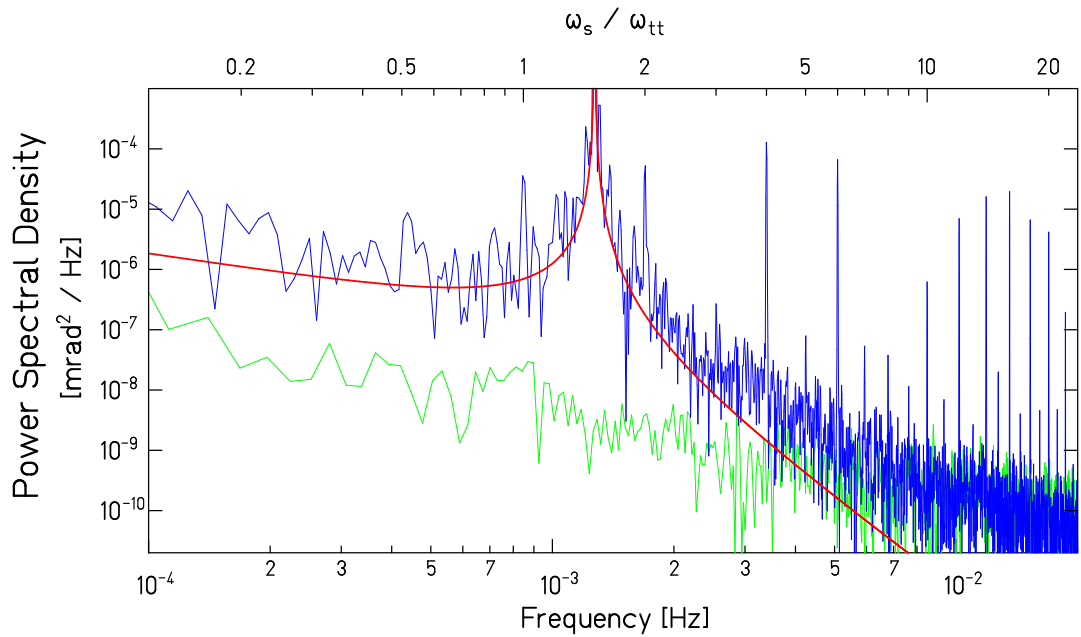}
\end{center}
\caption{[Colour online] Power spectral density of the twist signal. The upper [blue] histogram shows WEP data taken with $\omega_{\rm tt}/\omega_0=2/3$. The 
curve is the thermal noise predicted by equation \ref{eq: noise} for a room-temperature oscillator with $Q=6000$. The peaks at integer multiples of $\omega_{\rm tt}$ arise from reproduceable variations in $\omega_{\rm tt}$ (see equation \ref{eq: false effect}). The small peak at $\omega_{\rm tt}/2$ is caused by the turntable leveling system that recomputed the tilt every two turntable rotations\cite{he:08}.  The lower [green] histogram  displays data taken with the turntable stationary and the pendulum resting on a support to show the readout noise.  The low-frequency readout noise is ascribed to thermal fluctuations.}

\label{fig:noise}
\end{figure}

\begin{figure}[t]
\begin{center}
\includegraphics[width=0.6\columnwidth]{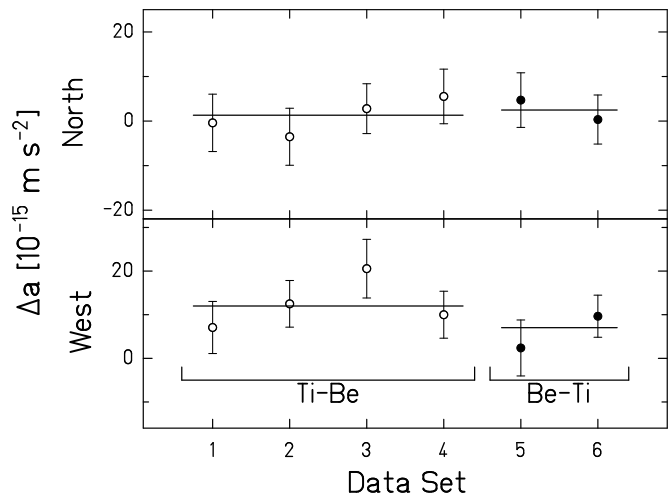}
\end{center}
\caption{ Data collected in the Ti-Be (first 4 runs) and Be-Ti (last 2 runs) configurations of the pendulum. The final result is in the difference between the means of the two configurations (shown as solid lines).}
\label{fig:data}
\end{figure}

Figure \ref{fig:noise} shows the power spectral density of the observed twist signal and demonstrates that the instrument operates close to the thermal limit.   
Figure \ref{fig:data} summarizes the Be-Ti composition dipole measurements. Each data point represents about two weeks of data, with daily reversals of the pendulum orientation with respect to the turntable rotor. A linear drift was removed to correct for slow environmental variations (the drift correction was insignificant compared to the statistical errors). The difference in the mean values for each configuration contains the signal. The offset from zero is due to systematic effects that follow the orientation of the pendulum frame. Approximately 75 days of data were collected using the Be-Ti test bodies and 110 days using the Be-Al test bodies. Systematic investigations were performed each time the vacuum system was pumped out and then repeated after the measurements were completed to ensure that the systematic effects had not changed.
%
%

Several environmental conditions are known to produce effects that can mimic a WEP-violating signal. Tilts of the rotation axis with respect to local vertical, coupling of the pendulum to gravity gradients, temperature fluctuations and gradients and magnetic fields all produce such effects. 
The systematic errors associated with these effects were measured following the strategy described in detail in \cite{su:94,he:08,sc:08}. Each ``driving term'' was deliberately exaggerated and its effect on the WEP-violating signal was measured; this signal was then scaled to the driving term observed in the actual WEP data.
Gravity gradients were measured with a specially designed gradiometer
pendulum that could be configured to give sensitivity to a particular
mulitpole component of the gradient; this information had been used to design the gradient compensators shown in figure~\ref{fig:apparatus}. Systematic errors from gravity gradients were measured by 
rotating the compensators by $180^{\circ}$ about the vertical axis,
so that instead of canceling the ambient gradient they effectively doubled it. The ratio of the twist signals with the WEP and gradiometer pendulums in the two compensator positions 
determined the effects of gravity gradients on the WEP pendulum;
this was used to correct the WEP signal.

It is well known that
small tilts of the apparatus induce a twist in the fibre because of
tiny asymmetries in the upper fibre attachment point.
Dual-axis electronic tilt sensors placed above the upper
attachment of the fibre and beneath the pendulum (see figure~\ref{fig:apparatus}) measured the turntable tilt. A feedback loop locked the
turntable rotation axis with a precision of a few nanoradians to local vertical as determined by the upper tilt sensor.  However, the lower tilt sensor revealed that the direction of local vertical at the pendulum position 
differed from that at the upper sensor by $\sim 50~\mbox{nrad}$. Since the fibre axis is determined by local level at the pendulum site, corrections were needed to account for this gradient in the down direction.  The tilt-induced twist was measured by purposely tilting the apparatus by a measured
amount. The resulting feed-through of a small tilt into pendulum twist
was typically around $5\%$, but varied from mirror to mirror.  Corrections for tilt
were applied to obtain the final result.

Temperature gradients and magnetic effects were primarily minimized by multi-stage passive shielding. The magnetic systematic uncertainty was found by removing the outermost mu-metal shield (which normally reduced the ambient laboratory field to $\approx 2.5\times 10^{-6}$ T) and measuring the effect on the twist signal when a strong permanent magnet was placed outside the vacuum vessel. In the absence of any shielding the magnet's field at the pendulum would have been $\approx1.7 \times 10^{-4}$~T. Data were taken with both the north and south poles pointing toward the pendulum. The pendulum twist did not significantly change when the magnet orientation was reversed. The magnetic systematic error was computed by scaling this upper limit on the twist change by the ratio of the normal to enhanced fields inside the outermost shield.
%
%
\begin{table}[b]
\caption{\label{tab:systematics}Error budget for the lab-fixed Be-Ti differential accelerations. Corrections were applied for gravitational gradients and tilt, only upper limits were obtained on the magnetic and temperature effects. All uncertainties are $1\,\sigma$.} 
\begin{indented}
\item[]\begin{tabular}{l rrrrrrrr}
\br
Uncertainty source        & \multicolumn{4}{c}{$\Delta a_{\rm N,Be-Ti} ~(10^{-15}$~ m s$^{-2})$} & \multicolumn{4}{c}{$\Delta a_{\rm W,Be-Ti} ~(10^{-15}$~ m s$^{-2})$} \\
\mr
Statistical							&	&	& $3.3\pm2.5$	& & & &	$-2.4\pm2.4$	&	\\
Gravity gradients				&	&	& $1.6\pm0.2$	& & & &	$0.3\pm1.7$		&	\\
Tilt										&	&	&	$1.2\pm0.6$	& & & &	$-0.2\pm0.7$	&	\\
Magnetic								&	&	&	$0\pm0.3$		& & & &	$0\pm0.3$			&	\\
Temperature gradients		&	&	&	$0\pm1.7$		& & & &	$0\pm1.7$			&	\\
\br
\end{tabular}
\end{indented}
\end{table}

The effect of temperature gradients was measured by placing large
temperature-controlled copper plates next to the apparatus and
measuring the pendulum signal as a function of the applied temperature
gradient. Temperature gradients of up to 15 K/m were
applied, while in normal operation the apparatus saw a gradient of $\sim 44\:$mK/m.
The maximum twist signal change in the temperature test was scaled to temperature gradients seen in normal data and assigned equally to systematic uncertainties in the north and west signals.

Table \ref{tab:systematics} summarizes the lab-fixed systematic effects in the Be-Ti measurement. When astronomical objects were viewed as the attractors, their additional signal modulation reduced the systematic uncertainties so that those results were dominated by the statistical uncertainty in contrast to the lab-fixed results where the statistical and systematic uncertainties were comparable.

The basic results from the E\"ot-Wash Be-Ti and Be-Al WEP tests are summarized in table \ref{tab:results}.
%
%
\begin{table}
\caption{\label{tab:results}Differential accelerations in the lab-fixed frame ($\Delta a_{\rm N}$ and $\Delta a_{\rm W})$ and toward the sun and galactic center ($\Delta a_{\odot}$ and $\Delta a_{\rm g}$). The E\"otv\"os parameters, 
$\eta_{\oplus}$, $\eta_{\odot}$ and $\eta_{\rm DM}$,were calculated using  the horizontal gravitational accelerations of earth, sun and galactic dark matter, 0.0168 m/s$^2$, $5.9\times 10^{-3}$ m/s$^2$ and $5\times 10^{-11}$ m/s$^2$ \cite{st:93}, respectively.
Uncertainties are $1\,\sigma$ with systematic and statistical uncertainties added in quadrature.}
\begin{indented}
\item[]\begin{tabular}{ll r rrr rrr }
\br
      					&	& & \multicolumn{3}{c}{Be-Ti} & \multicolumn{3}{c}{Be-Al} \\
\mr      						
$\Delta a_{\rm N}$ &  $(10^{-15}$~ m s$^{-2})$ & & &  $0.6\pm3.1$  & & & $-1.2\pm2.2$  & \\
$\Delta a_{\rm W}$ &  $(10^{-15}$~ m s$^{-2})$ & & &	$-2.5\pm3.5$ & & & $0.2\pm2.4$   & \\
$\Delta a_{\odot}$ &  $(10^{-15}$~ m s$^{-2})$ & & &	$-1.8\pm2.8$ & & & $-3.1\pm2.4$  & \\
$\Delta a_{\rm g}$ &  $(10^{-15}$~ m s$^{-2})$ & & &	$-2.1\pm3.1$ & & & $-1.2\pm2.6$  & \\
$\eta_{\oplus}$ 	 &  $(10^{-13})$ 						 & & &  $0.3\pm1.8$  & & & $-0.7\pm1.3$  & \\
$\eta_{\odot}$ 	 	 &  $(10^{-13})$ 						 & & &  $-3.1\pm4.7$  & & & $-5.2\pm4.0$ & \\
$\eta_{\rm DM}$ 			 &  $(10^{-5})$ 						 & & &  $-4.2\pm6.2$  & & & $-2.4\pm5.2$ & \\
\br
\end{tabular}
\end{indented}
\end{table}
\section{Some implications of the results for new Yukawa interactions}
\subsection{Results}
\begin{figure}[t]
\begin{center}
\includegraphics[width=1.0\columnwidth]{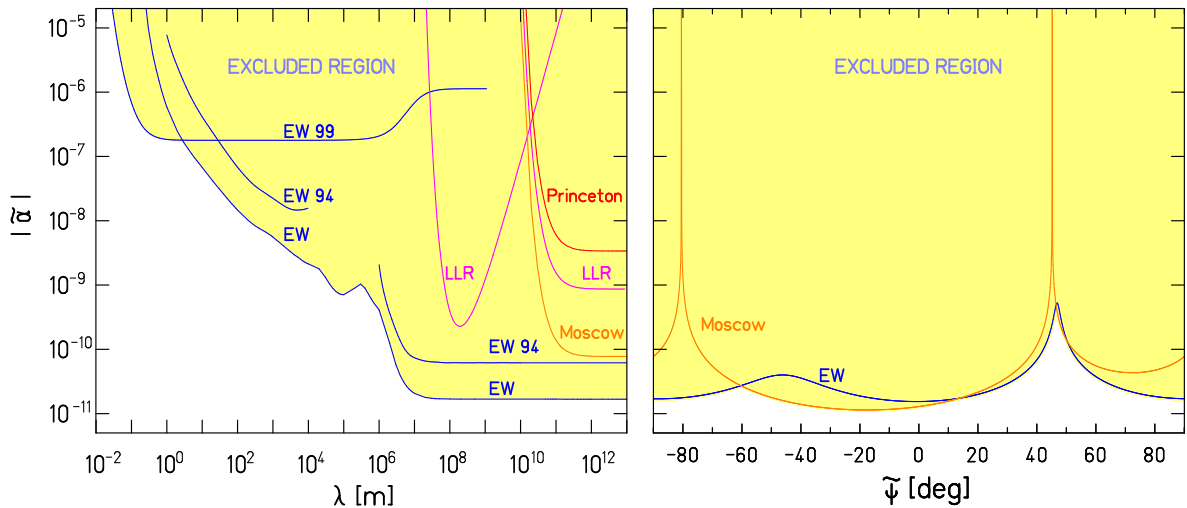}
\end{center}
\caption{The left panel shows $95\%$ CL upper bounds on the strength of a vector Yukawa interaction 
coupled to $\tilde{q} = B - L$. The curve labeled EW shows the limit extracted from the lab-fixed results in table \ref{tab:results}. Curves labeled Princeton, Moscow, EW94, and EW99 are extracted from \cite{ro:64}, \cite{br:71}, \cite{su:94} and  \cite{sm:00}, respectively. The two LLR constraints are derived from lunar laser-ranging results\cite{tu:07} for the earth-moon differential acceleration toward the sun (right curve) and the inverse-square law violation obtained from anomalous precession of the lunar orbit (left curve). The right panel shows how the constraints on an infinite-range interaction depend on $\tilde{\psi}$ the parameter that describes the interaction charge. The combined E\"ot-Wash result from Be-Ti and Be-Al attracted toward the earth and toward the sun is indicated by EW. The Moscow result\cite{br:71} used a Al-Pt dipole attracted to the sun; the left pole arises when the sun's charge is zero, while the pole on the right occurs where the charge difference of the test bodies vanishes.}
\label{fig:alphalambda}
\end{figure}

The properties of our terrestrial attractor allow the lab-fixed Be-Ti and Be-Al results in table  \ref{tab:results} to constrain exotic Yukawa interactions with ranges down to 1 m. A torsion balance located on a flat, level region would have essentially no sensitivity for forces with $\lambda \le r_{\rm earth}$ (see \cite{ad:90}). However, the E\"ot-Wash laboratory is located on a hillside above a deep lake, with the pendulum only 0.75 m from a wall excavated from the hillside. The complex regional topography plus details of the laboratory environment enormously enhance the sensitivity for Yukawa interactions with $\lambda < 10^7$ m. But determining the sensitivity for such forces is a challenging undertaking because one needs to compute  the {\em horizontal} component of a Yukawa force from a complicated object. We estimated this strength using geophysical models extending from the detailed local topography to regional geology\cite{fi:05,st:07} to the gross structure of the earth\cite{la:11,st:00,dz:81}. 
The left panel in figure \ref{fig:alphalambda} shows $95\%$ CL limits on the magnitude of the Yukawa strength $\tilde{\alpha}$ as a function of range $\lambda$ assuming a charge  $\tilde{q} = N= B-L$,
where $B$ and $L$ are the baryon and lepton numbers, respectively. This is a particularly interesting charge because $B-L$ is conserved in grand unified theories. The bump in Figure~\ref{fig:alphalambda} at $\lambda \sim 10^5$ m comes from an east-west density
asymmetry in the subduction zone for the Juan de Fuca plate. As $\lambda$ increases beyond 60 km, which is
approximately the depth of the subduction zone, 
the supporting mantle 
quickly reduces the asymmetry to maintain
hydrostatic equilibrium.
%
%
%
Constraints on vector interactions coupled to other charges can be inferred from the right panel in figure \ref{fig:alphalambda}, which  displays $95\%$ CL limits on $|\tilde{\alpha}|$ as a function of $\tilde{\psi}$ for an infinite-ranged interaction. Since any single pair of test bodies (or source) has a value of $\tilde{\psi}$ for which its charge difference (or charge) vanishes, two different pairs of test bodies and two different sources must be used to  obtain limits for all values of $\tilde{\psi}$. 

Figure \ref{fig:dilaton} shows an example of WEP bounds on scalar interactions, the Donoghue-Damour\cite{do:10a} scenario for WEP violation by massless dilatons. Their predicted WEP-violating effects are dominated by couplings to the average light quark mass and the electromagnetic field strength via the ``dilaton coefficients'' $D_{\hat{m}}$ and $D_{e}$, respectively.  Our 95\% CL limits in the $D_{\hat{m}}$-$D_{e}$ parameter space demonstrate that the effects of a massless dilaton must be suppressed by a factor of at least $\sim 10^{10}$. (The  individual 95~\%CL constraints on $D_{\hat{m}}$ and  $D_{e}$ are $(-0.3\pm 3.2)\times 10^{-10}$ and $(+1.7\pm 10.3)\times 10^{-10}$, respectively.) This suggests that the dilaton must have a finite mass so that its  short-range force was not detected in WEP experiments. In this case, inverse-square law tests, which probe the dominant composition-independent coupling to the gluon strength, set a conservative lower limit of 3.5 meV on the dilaton mass\cite{ad:07,ka:00}. This lower limit becomes 13 meV in the standard model if the string scale is set to the Planck scale{\cite{ka:00}.
%
%
\begin{figure}[ht]
\begin{center}
\includegraphics[width=0.7\columnwidth]{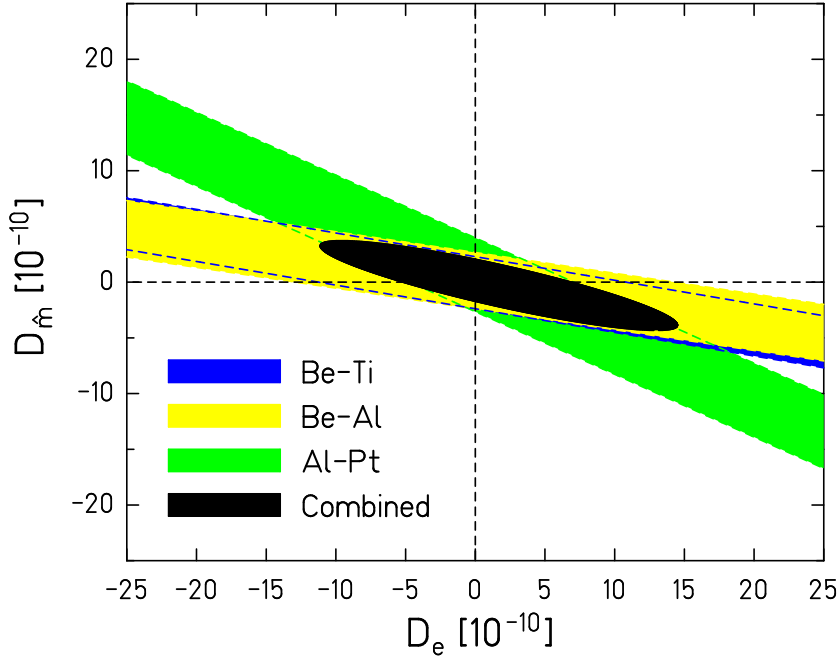}
\end{center}
\caption{[Colour online] 95\% CL constraints on couplings of a long-range dilaton field. We use our Be-Ti and Be-Al results plus the Moscow Al-Pt limit\cite{br:71} to constrain the dominant dilaton couplings $D_e$ and $D_{\hat{m}}$ in the Donoghue-Damour framework. For standard dilaton couplings these coeffcients would be of order unity.}
\label{fig:dilaton}
\end{figure}
\subsection{Some implications of the results}
\label{subsec:implications}
The impressive recent technical progress\cite{al:11,en:10} toward trapping antihydrogen, $\bar{\rm H}$, has revived interest in probing the gravitational properties of antimatter by testing the suggestion that antimatter could fall with an acceleration perceptibly different from $g$\cite{ni:91}.
It is worth asking how plausible this is, especially considering the extraordinary technical difficulties involved in measuring the freefall acceleration of antihydrogen. In field theory terms, if antihydrogen were to fall with an acceleration different from hydrogen it could occur if and only if there were a vector interaction that coupled to $Z$. But the WEP results summarized above set extremely strong upper limits on such vector interactions. 
To be explicit, what should one expect if one could do a hydrogen-antihydrogen freefall comparison at the location of the E\"ot-Wash WEP torsion balance? To answer this, we used our geophysical earth model to calculate the ratio, as a function of $\lambda$, of the vertical to horizontal Yukawa forces at our site. Then, using our constraint on $\tilde{\psi}= 0$ vector interactions, we computed the upper bound on the vertical component of $\Delta a_{\rm \bar{H}-H}/g$. The results, shown in figure~\ref{fig:antih}, indicate that any anomalous gravitational acceleration must be extremely small, well below the sensitivity of current technology. One might object that our $\tilde{\psi}=0$ assumption is unwarranted because it assumes that antineutrons and neutrons should fall with identical accelerations. Indeed it is, so we also computed the upper bound on $\Delta a_{\rm \bar{H}-H}/g$ for the values of $\tilde{\psi}$ 
that gave the weakest constraint at each value of $\lambda$; the results are also shown in figure~\ref{fig:antih}. Had we plotted the corresponding constraints on $\Delta a_{\rm \bar{n}-n}/g$ 
, a quantity that would be essentially impossible to measure directly, the results would be essentially the same as the hydrogen-antihydrogen bounds in figure~\ref{fig:antih}. 

It has been argued\cite{ni:91} that the existence of a scalar field could invalidate these arguments; the scalar field would have no effect on $\Delta a_{\rm \bar{H}-H}$ because particles and antiparticles have the same scalar charge, but it would contribute to the differential accelerations of the E\"ot-Wash test bodies. In fact, because scalar forces between like particles are attractive, a scalar interaction would tend to cancel a vector force. But this cancelation must be unreasonably precise to give null results in WEP tests with 9 different materials (ranging from Be to Pb) falling toward 3 different attractors\cite{br:71,su:94,sm:00,wa:12}. Suppose the scalar charges of the materials used for these tests differed  by merely 0.1\%  from the vector charges in equation \ref{eq: charge}; the upper limit on $\Delta a_{\rm \bar{H}-H}/g$ from a long-range vector field would still be about 1 part in $10^6$.  Reference \cite{al:09} gives a detailed discussion of the impossibility of nearly perfect scalar-vector cancellation. Of course, our arguments rely on the $CPT$ theorem that, to our knowledge, has not been tested for gravity. But consider how strange it would be if, as is occasionally suggested, antimatter fell up rather than down. In that case a particle that is its own antiparticle (such as the photon or $\pi^0$) would not fall. This is excluded by many observations. 
%
%
\begin{figure}[t]
\begin{center}
\includegraphics[width=0.7\columnwidth]{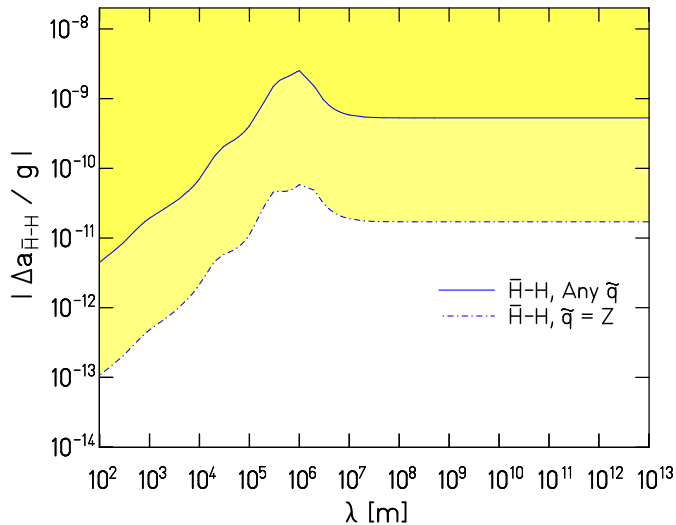}
\end{center}
\caption{ $95\%$ CL upper bounds on $|\Delta a_{H \bar{H}}| / g $, the fractional difference in the freefall accelerations of hydrogen and antihydrogen. The region above the solid line is excluded for all values of $\tilde{q}$. The region above the dashed line is excluded if we assume that $\tilde{q}=Z$. The corresponding constraints for neutrons and antineutrons are almost the same as those for hydrogen and antihydrogen.}
\label{fig:antih}
\end{figure}

WEP results also provide a laboratory test of the common assumption that gravitation is the only long-range force between dark and luminous matter. Because almost all of the usual conclusions about dark matter rely on this assumption, finding laboratory support for the idea has real value. The acceleration vector toward 
the galaxy's dark matter passes through our instrument's plane of maximum sensitivity and has an estimated magnitude of 
$5\times 10^{-11}$ m s$^{-2}$\cite{st:93}. This acceleration can be separated into gravitational and non-gravitational components $a_{\rm DM} = a_{\rm DM}^{\rm g} + a_{\rm DM}^{\rm ng}$. We assume that any non-gravitational interaction with dark matter violates the WEP and search for differential accelerations $\delta a_{\rm DM}^{\rm ng}$ for a series of test-body pairs. From these results we can deduce, for a given material, the magnitude of  $a_{\rm DM}^{\rm ng}$ as a function of $\tilde{\psi}$, the parameter describing the test-body charges (see ~\cite{su:94}). We combine the Be-Al and Be-Ti  galactic attractor results from table \ref{tab:results} with the $3\times 10^{-16}$ m s$^{-2}$ upper bound on the earth-moon differential acceleration  toward the galactic center\cite{no:95} extracted from lunar laser-ranging (LLR) data 
 to obtain an upper bound on the contribution of non-gravitational forces to the galactic dark-matter acceleration of neutral hydrogen shown in 
figure \ref{fig:darkmatter}. 
Extraordinarily, for any value of $\tilde{\psi}$, the acceleration of hydrogen due to non-gravitational interactions with dark matter must be less than about $5\%$ of the total acceleration. The bounds in figure \ref{fig:darkmatter} apply to
any interaction whose WEP violation is approximately described by equation~\ref{eq: charge}. For example, a very similar upper bound would arise for a dilaton-like scalar field. The WEP-violating component of the dilaton coupling to the gluon strength is about 0.3\%\cite{ka:00} which is very similar to the 0.25\% difference in $B/\mu$ values of Be and Ti.
%
%
\begin{figure}[ht]
\begin{center}
\includegraphics[width=0.7\columnwidth]{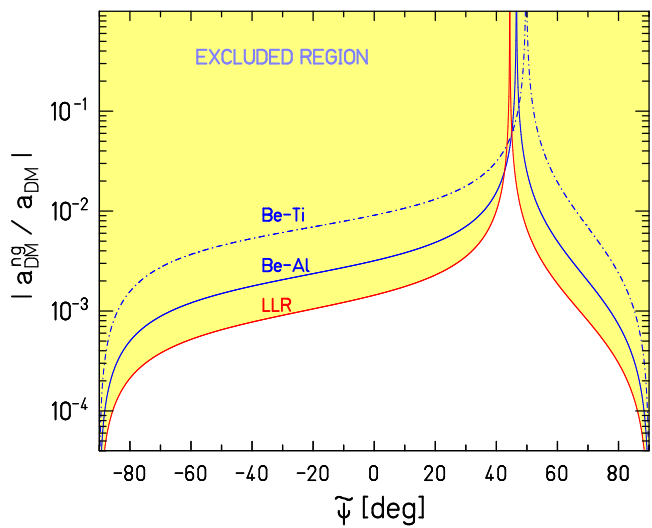}
\end{center}
\caption{Inferred $95\%$ CL limits on the ratio of non-gravitational acceleration of neutral hydrogen to the total acceleration toward galactic dark matter. The ratio vanishes at $\tilde{\psi}=\pm 90^{\circ}$ where the charge of hydrogen is zero. The weakest bound of about 5\% occurs near $\tilde{\psi} = 45^{\circ}$ where $\tilde{q} \propto B$.}
\label{fig:darkmatter}
\end{figure}
\section{Prospects for further improvements}
EP tests may be characterized by the sensitivity figure
\begin{equation}
S=\frac{\Delta_{\rm TB}}{\delta(\Delta_a)}=\frac{\Delta({\tilde q}/\mu)}{\delta(\Delta_a)}~,
\label{eq: FOM}
\end{equation}
where $\delta(\Delta_a)$ is the uncertainty in the measured test-body differential acceleration $\Delta_a$ and $\Delta_{\rm TB}$ measures the difference in test body compositions; we characterize $\Delta_{\rm TB}$ in terms of the differential charge-to-mass ratio given in equation \ref{eq: charge}.
The E\"ot-Wash group is designing a next-generation WEP experiment with the goal of improving the precision of our existing test\cite{sc:08,ad:09} by an order of magnitude. The existing result was limited by statistical errors (predominantly thermal noise from anelastic losses in the suspension fibre) and by systematic errors from time-varying gravity gradients (see ~\cite{su:94} for a discussion of this problem). The new 
experiment aims for an order of magnitude improvement in $S$ by increasing the numerator in equation \ref{eq: FOM} with a new pendulum having a higher composition contrast,
$\Delta_{\rm TB}$, and by decreasing the denominator with reduced thermal noise and systematic errors from time-varying gravity gradients. 

A new pendulum with neutron-rich beryllium and proton-rich polyethylene test bodies will
increase $\Delta({\tilde q}/\mu)$ by about a factor of 10 (see ~\cite{ad:09}).
The thermal noise power in a torsional oscillator with torsional constant $\kappa$ and quality factor $Q$ is proportional to $\kappa/Q$ (see equation \ref{eq: noise}).  The next generation of 
E\"ot-Wash WEP experiments will employ silica fibres because 
fused silica  provides a much lower $\kappa / Q$ ratio than tungsten. The E\"ot-Wash WEP experimenters had been concerned that silica's insulating property would be a problem, but Stefano Vitale's group in Trento used fused-silica fibres prepared by the Glasgow group and found that problems from pendulum charging are perfectly manageable (less than 1 elementary charge per second)\cite{ca:09}. They obtained $Q$s as high as $\approx 740,000$ 
but the noise power at mHz frequencies was lower than that of a tungsten fibre by only a factor of 6. The reason for this is not yet understood, so that there is a possibility of yet further improvements in the statistical errors. The problem of time-varying gravity gradients will be addressed by continuous measurements of the leading term in the gradient.
\ack
The work we describe would not have been possible without the ideas and inventivess of many members of the E\"otWash group, 
particularly Blayne Heckel.
This work was primarily supported by NSF Grant PHY0969199 and supplemented by funds from DOE support for the University of Washington Center for Nuclear and Particle Astrophysics.
\section*{References}

\bibliographystyle{iopart-num}
\bibliography{WEP}

\providecommand{\newblock}{}
\begin{thebibliography}{10}
\expandafter\ifx\csname url\endcsname\relax
  \def\url#1{{\tt #1}}\fi
\expandafter\ifx\csname urlprefix\endcsname\relax\def\urlprefix{URL }\fi
\providecommand{\eprint}[2][]{\url{#2}}

\bibitem{ro:64}
Roll P~G, Krotkov R and Dicke R~H 1964 {\em Ann. Phys., NY\/} {\bf 26} 442--517

\bibitem{br:71}
Braginsky V~B and Panov V~I 1971 {\em ZhEksp. Teor. Fiz.\/} {\bf 61} 873--1272

\bibitem{fi:86}
Fischbach E, Sudarsky D, Szafer A, Talmadge C and Aronson S~H 1986 {\em Phys.
  Rev. Lett.\/} {\bf 56} 3--6

\bibitem{ka:00}
Kaplan D~B and Wise M~B 2000 {\em J. High Energy Phys.\/} {\bf JHEP08} 037
  (\textit{Preprint} \eprint{hep-ph/0008116})

\bibitem{do:10}
Damour T and Donoghue J~F 2010 {\em Phys. Rev.\/} D {\bf 82} 084033
  (\textit{Preprint} \eprint{1007.2792})

\bibitem{do:10a}
Damour T and Donoghue J~F 2010 {\em Class. Quantum Grav.\/} {\bf 27} 202001
  (\textit{Preprint} \eprint{1007.2790})

\bibitem{ad:90}
Adelberger E~G, Stubbs C~W, Heckel B~R, Su Y, Swanson H~E, Smith G~L, Gundlach
  J~H and Rogers W~F 1990 {\em Phys. Rev.\/} D {\bf 42} 3267--3292

\bibitem{su:94}
Su Y, Heckel B~R, Adelberger E~G, Gundlach J~H, Harris M, Smith G~L and Swanson
  H~E 1994 {\em Phys. Rev.\/} D {\bf 50} 3614--3636

\bibitem{sa:90}
Saulson P~R 1990 {\em Phys. Rev.\/} D {\bf 42} 2437--2445

\bibitem{pe:11}
Pegna R, Nobili A~M, Shao M, Turyshev S~G, Catastini G, Anselmi A, Spero R,
  Doravari S, Comandi G~L and De~Michele A 2011 {\em Phys. Rev. Lett.\/} {\bf
  107} 200801

\bibitem{sm:00}
Smith G~L, Hoyle C~D, Gundlach J~H, Adelberger E~G, Heckel B~R and Swanson H~E
  1999 {\em Phys. Rev.\/} D {\bf 61} 022001

\bibitem{he:08}
Heckel B~R, Adelberger E~G, Cramer C~E, Cook T~S, Schlamminger S and Schmidt U
  2008 {\em Phys. Rev.\/} D {\bf 78} 092006 (\textit{Preprint}
  \eprint{0808.2673})

\bibitem{sc:08}
Schlamminger S, Choi K~Y, Wagner T~A, Gundlach J~H and Adelberger E~G 2008 {\em
  Phys. Rev. Lett.\/} {\bf 100} 041101 (\textit{Preprint} \eprint{0712.0607})

\bibitem{wa:12}
Wagner T~A 2012 Ph.D. thesis {U}niversity of {W}ashington unpublished

\bibitem{ho:04}
Hoyle C~D, Kapner D~J, Heckel B~R, Adelberger E~G, Gundlach J~H, Schmidt U and
  Swanson H~E 2004 {\em Phys. Rev.\/} D {\bf 70} 042004 (\textit{Preprint}
  \eprint{hep-ph/0405262})

\bibitem{st:93}
Stubbs C~W 1993 {\em Phys. Rev. Lett.\/} {\bf 70} 119--122

\bibitem{fi:05}
Finlayson D 2005 Combined bathymetry and topography of the {Puget Lowland},
  {Washington} state University of Washington website
  http://www.ocean.washington.edu/data/pugetsound/

\bibitem{st:07}
Stephenson W~J 2007 Velocity and density models incorporating the {Cascadia}
  subduction zone for {3D} earthquake ground motion simulations, version 1.3
  Open-File Report 2007-1348 U.S. Geological Survey, Earthquake Hazards Ground
  Motion Investigations

\bibitem{la:11}
Laske G, Dziewonski A and Masters G 2011 Reference earth model website
  http://igppweb.ucsd.edu/$\sim$gabi/rem.html

\bibitem{st:00}
Steinberger B 2000 {\em Phys. Earth Planet. Inter.\/} {\bf 118} 241--257

\bibitem{dz:81}
Dziewonski A~M and Anderson D~L 1981 {\em Phys. Earth Planet. Inter.\/} {\bf
  25} 297--356

\bibitem{tu:07}
Turyshev S~G and Williams J~G 2007 {\em Int. J. Mod. Phys.\/} D {\bf 16}
  2165--2179 (\textit{Preprint} \eprint{gr-qc/0611095})

\bibitem{ad:07}
Adelberger E~G, Heckel B~R, Hoedl S, Hoyle C~D, Kapner D~J and Upadhye A 2007
  {\em Phys. Rev. Lett.\/} {\bf 98} 131104

\bibitem{al:11}
Andresen G~B {\em et~al.\/} (The {ALPHA} Collaboration) 2011 {\em Nature
  Phys.\/} {\bf 7} 558--564 (\textit{Preprint} \eprint{1104.4982})

\bibitem{en:10}
Enomoto Y {\em et~al.\/} 2010 {\em Phys. Rev. Lett.\/} {\bf 105} 243401

\bibitem{ni:91}
Nieto M~M and Goldman T 1991 {\em Phys. Rept.\/} {\bf 205} 221--281

\bibitem{al:09}
Alves D~S~M, Jankowiak M and Saraswat P 2009  (\textit{Preprint}
  \eprint{0907.4110})

\bibitem{no:95}
Nordtvedt K~L, M\"uller J and Soffel M 1995 {\em Astron. Astrophys.\/} {\bf
  293} L73--L74

\bibitem{ad:09}
Adelberger E~G, Gundlach J~H, Heckel B~R, Hoedl S and Schlamminger S 2009 {\em
  Prog. Part. Nucl. Phys.\/} {\bf 62} 102--134

\bibitem{ca:09}
Cavalleri A {\em et~al.\/} 2009 {\em Class. Quantum Grav.\/} {\bf 26} 094017

\end{thebibliography}
\end{document}